\documentclass[aps,12pt,nofootinbib]{revtex4-1}

\usepackage{epsfig}
\usepackage{graphicx,graphics}
\usepackage{amsmath}
\usepackage{amsfonts}
\usepackage{amssymb}

\begin{document}

\title{The asymmetric simple exclusion process on chains with a shortcut revisited}

\author{Nadezhda Bunzarova$^{1,2}$, Nina Pesheva$^2$, and Jordan Brankov$^{1,2}$}

\affiliation{$^1$ Bogoliubov Laboratory of Theoretical Physics, Joint Institute
for Nuclear Research, 141980 Dubna, Russian Federation\\
$^2$ Institute of Mechanics, Bulgarian Academy of Sciences, 1113 Sofia, Bulgaria}

\begin{abstract}
We consider the asymmetric simple exclusion process (TASEP) on open network consisting of three
consecutively coupled macroscopic chain segments with a shortcut
between the tail of the first segment and the head of the third one. The model was introduced
by Y.-M. Yuan et al [J. Phys. A 40, 12351 (2007)] to describe directed motion of molecular motors along
filaments. We report here unexpected results in the case of maximum current through the network which
revise the previous findings.
Our theoretical analysis, based on the effective rates approximation, shows that the second (shunted)
segment can exist in both low-density and high-density phases, as well as in the coexistence (shock) phase.
Numerical simulations demonstrate that it is the last option that takes place - the local density distribution
and the nearest-neighbor correlations in the middle chain correspond to a shock phase with completely
delocalized domain wall. Surprisingly, the main quantitative parameters of that shock phase are governed
by a real root of a cubic equation the coefficients of which simply depend on the probability of choosing
the shortcut. The unexpected conclusion is that a shortcut in the bulk of a single lane always creates
traffic jams.

\vspace{0.5cm}

\noindent {\it Pacs:} {02.50.Ey, 05.60.-k, 05.70.Ln}

\vspace{0.5cm}


\end{abstract}

\maketitle

{\it Introduction.} --- The asymmetric simple exclusion process (TASEP) is one of the paradigmatic models
for understanding the rich world of non-equilibrium phenomena. Devised to model kinetics of protein synthesis
\cite{MGP68}, it has found a number of applications to vehicular traffic flow \cite{vehtraf}, biological
transport \cite{biotraf}, one-dimensional surface growth \cite{surf}, forced motion of colloids in
narrow channels \cite{zeo}, spintronics \cite{spin}, transport of 'data packets' in the Internet
\cite{Inter}, current through chains of quantum dots \cite{KO10}, to mention some.

Novel features of the TASEP have been found on networks consisting
of coupled linear chains with nontrivial geometry. In the approach
advanced in our work \cite{BPB04} each macroscopic segment $s$ of
the network is considered in a stationary phase determined by its
effective injection $\alpha_s^*$ and ejection $\beta_s^*$ rates.
At that, exact in the thermodynamic limit results for the density
profile are incorporated. The only molecular field type
approximation used consists in the neglect of correlations between
different chain segments. This allows one to treat the coupling
between each two connected segments as coupling to reservoirs with
certain effective rates. The possible phase structures of the
whole network are obtained as solutions of the resulting set of
equations for the effective rates that follow from continuity of
current. The importance of our approach for modeling complex
biological transport phenomena was pointed out by Pronina and
Kolomeisky \cite{PK05}. This method became very popular and was
used in a number of studies of TASEP and its generalizations on
networks with different geometries, e.g., with junctions,
bifurcations, intersections, interacting lanes \cite{eff}.
Finite-size effects on the density profile due to shifting the
position of the double-chain section from the middle of the linear
network were studied too \cite{PB13}.

Here we consider the TASEP on open chain with a shortcut in the
bulk, introduced as 'model A' in \cite{YJWHW07}. The current
through the shortcut is proportional to a probability $q$. It is
convenient to consider the system as composed of three
consecutively connected macroscopic chain segments and a shortcut
between the tail of the first segment and the head of the third
one. In principle, the effect of a shortcut can easily be
understood: the decrease in the current through the shunted part
(second segment) of the original chain leads to a sharp change of
the particle density in the latter. If the chain without a
shortcut ($q=0$) is in the low-density (LD) phase, its bulk
density $\rho_{\rm bulk}^{\rm LD}<1/2$ supports a current $J
=\rho_{\rm bulk}^{\rm LD}(1-\rho_{\rm bulk}^{\rm LD}) < 1/4$. The
shortcut takes a part $J^{\rm sc}>0$ of that current away from the
second segment, hence the current $J^{(2)} = J- J^{\rm sc}$ has to
be supported by still less bulk density $\rho_{\rm bulk}^{(2)}
<\rho_{\rm bulk}^{\rm LD}$ in that segment. Similarly, when the
initial chain is in the high-density (HD) phase with $\rho_{\rm
bulk}^{\rm HD}> 1/2$, the drop in the current through the second
segment, caused by the shortcut, leads to a still higher bulk
density in that segment, $\rho_{\rm bulk}^{(2)}> \rho_{\rm
bulk}^{\rm HD}$. In these cases all the three segments remain in
the same phase, though with different density of the middle one.

Not so clear, however, is the situation when the initial chain is
in the maximum current (MC) phase with $\rho_{\rm bulk}^{\rm MC}=
1/2$. Now the drop in the current through the shunted segment of
the network can be compensated equally well by decrease or
increase in its bulk density. Then, the middle segment is forced
either in low-density, or in high-density phase, which may lead
also to coexistence of LD phase on the left-hand side with HD
phase on the right-hand side. This phase structure is additionally
favored by the downward (upward) bend in the density profile of
the first (third) segment in the maximum current phase. In the
case of open system with variable total number of particles the
coexisting phases are likely to be separated by a completely
delocalized domain wall. Such was the situation observed in each
of the equivalent segments in a double-chain section
incorporated in the middle of a long linear chain
\cite{BPB04}. It seems plausible that the above
mechanism of influence of the shortcut on the phase state of the
shunted segment should be invariant with respect to the explicit
structure of the shortcut. In particular, one may consider a
shortcut in the form of an additional (shorter) chain connecting
the last site of the first segment to the first site of the third
one. Since the length of the shortcut is irrelevant, we can
include the case of parallel segments with equal length considered
in our work \cite{BPB04}.  However, the authors of \cite{YJWHW07}
have claimed that in the case of their 'model A', the shunted
middle segment can exist only in the high-density phase. This
contrast in the  conclusions motivated us to renew the study, both
analytically and numerically, of the model. The results may
have important implications for vehicular traffic flow control,
as well as for biological transport in living cells.

{\it Microscopic model.} --- Here we consider model A of a shortcut,
suggested in \cite{YJWHW07}, when both the injection $\alpha$ and
ejection $\beta$ rates at the open ends of the system are larger than 1/2,
so that the first and third segments are in the maximum current phase. The
shortcut is between the last site of the first segment, with occupation number $\tau_L^{\rm (1)}$,
and the first site of the third
segment, with occupation number $\tau_1^{\rm (3)}$. According the rules of the random-sequential
algorithm, when a particle at the last site of the first segment (with $\tau_L^{\rm (1)}=1$)
attempts to move, the particle may
jump along the main track to the first site of the second segment with rate $(1-q)(1-\tau_1^{\rm (2)})(1-\tau_1^{\rm (3)}) + (1-\tau_1^{\rm (2)})\tau_1^{\rm (3)}$, or take the
shortcut to the first site of the third segment with rate
$q(1-\tau_1^{\rm (3)})$, or stay immobile with rate $(1-q)\tau_1^{\rm (2)}(1-\tau_1^{\rm (3)}) +
\tau_1^{\rm (2)}\tau_1^{\rm (3)}$. These rules lead to exact expressions for the
stationary current through segment 2,
\begin{equation}
J^{\rm (2)} = (1-q)\langle \tau_L^{\rm (1)}(1-\tau_1^{\rm (2)})(1-\tau_1^{\rm (3)})\rangle +
\langle \tau_L^{\rm (1)}(1-\tau_1^{\rm (2)})\tau_1^{\rm (3)}\rangle = \langle \tau_L^{\rm (2)}(1-\tau_1^{\rm (3)})\rangle,
\label{ec2}
\end{equation}
and through the shortcut $J^{\rm sc} = q\langle \tau_L^{\rm (1)}(1-\tau_1^{\rm (3)})\rangle$,
$\quad 0\leq q \leq 1$.

{\it Theoretical analysis.} --- In the effective rates analysis
\cite{BPB04} one neglects the correlations between sites belonging
to different segments, so that the above expressions simplify to
\begin{equation}
J^{\rm (2)} = \rho_L^{\rm (1)}(1-\rho_1^{\rm (2)})[(1-q)(1-\rho_1^{\rm (3)}) +
 \rho_1^{\rm (3)}]= \rho_L^{\rm (2)}(1-\rho_1^{\rm (3)}),
\label{mfcsc}
\end{equation}
and $J^{\rm sc} = q\rho_L^{\rm (1)}(1-\rho_1^{\rm (3)})$,
where $\rho_i^{\rm (s)}=\langle \tau_i^{\rm (s)}\rangle$, ${\rm s}=1,2,3$, is the average value of
the occupation number $\tau_i^{\rm (s)}$ in a given stationary state. Within the above
approximation effective injection, $\alpha_{\rm s}^*$, and ejection, $\beta_{\rm s}^*$,
rates for segment ${\rm s}=1,2,3$ are introduced according to the rule \cite{BPB04} $
J^{\rm (s)} = \beta_{\rm s}^* \rho_L^{\rm (s)}= \alpha_{\rm s}^*(1-\rho_1^{\rm (s)})$,
with $\alpha_{\rm 1}^* =\alpha$ and $\beta_{\rm 3}^* = \beta$. Thus,
taking into account that $J^{\rm (1)} = J^{\rm (3)} = J^{\rm (2)}+ J^{\rm sc}$, one obtains
\begin{eqnarray}
\alpha_{\rm 1}^* &=&\alpha, \quad \beta_{\rm 1}^* = 1-\rho_1^{\rm (2)}
+q\rho_1^{\rm (2)}(1-\rho_1^{\rm (3)}), \label{1} \\
\alpha_{\rm 2}^*&=&\rho_L^{\rm (1)}[1-q(1-\rho_1^{\rm (3)})], \quad \beta_{\rm 2}^* =
1-\rho_1^{\rm (3)}, \label{2} \\
\alpha_{\rm 3}^*&=& \rho_L^{\rm (2)}+ q\rho_L^{\rm (1)}, \quad \beta_{\rm 3}^* =\beta.
\label{3}
\end{eqnarray}
Here we have taken into account that $\alpha_{\rm 3}^*$ comes from both the
expression for the current $J^{\rm sc}$
and the last one in (\ref{mfcsc}). Expressions (\ref{1})-(\ref{3}) for
the effective rates coincide exactly with equations (4) obtained
in \cite{YJWHW07}. However, the results of our analysis in the case when the
first and third segments are in the maximum current phase are essentially
different from those claimed in \cite{YJWHW07}.

We confine ourselves to the study of possible phase structures of
the type $(M,X,M)$, when the first and third segments are in the
maximum current phase $M$, and the second segment is in a
low-density phase ($X=L$), high-density one ($X=H$), or on the
coexistence line ($X=C$). Note that the case ($X=M$) is excluded,
since the presence of a shortcut ($J^{\rm sc}>0$) implies $J^{\rm
(2)} < J^{\rm (1)} = J^{\rm (3)} =1/4$. To check the consistency
of a given structure $(M,X,M)$ with the corresponding conditions
on the effective rates (\ref{1})-(\ref{3}), we make use of the
known, exact in the thermodynamic limit, values of the bulk
density $\rho_{\rm bulk}^{\rm (s)}$ and local densities
$\rho_1^{\rm (s)}$, $\rho_L^{\rm (s)}$, in dependance on the
thermodynamic phase of each segment $s=1,2,3$ \cite{D}. At that we
assume that all the segments have an equal, large enough length $L
\gg 1$. Thus, in all cases under consideration one has
\begin{eqnarray}
\rho_{\rm bulk}^{\rm (1)} = 1/2, \quad \rho_1^{\rm (1)} = 1-1/(4\alpha),
\quad \rho_L^{\rm (1)} = 1/(4\beta_{\rm 1}^*)\label{s1}\\
\rho_{\rm bulk}^{\rm (3)} = 1/2, \quad \rho_1^{\rm (3)} = 1-1/(4\alpha_{\rm 3}^*),
\quad \rho_L^{\rm (3)} = 1/(4\beta).\label{s3}
\end{eqnarray}
By inserting the expressions for $\rho_L^{\rm (1)}$ and $\rho_1^{\rm (3)}$ into Eq. (\ref{2}),
we obtain
\begin{equation}
\alpha_{\rm 2}^* =1/(4\beta_{\rm 1}^*)- q/(16\alpha_{\rm 3}^*\beta_{\rm 1}^*),
\quad \beta_{\rm 2}^* =1/(4\alpha_{\rm 3}^*),
\label{s2eff}
\end{equation}
and, from Eq. (\ref{mfcsc}), $J^{\rm sc} = 1/4 - J^{\rm (2)} = q/(16\beta_{\rm 1}^*\alpha_{\rm 3}^*)$.
Now we pass to the separate consideration of each of the possibilities $X=L,H,C$.

{\it Middle segment in the low-density phase.} --- In this case
\begin{equation}
\rho_{\rm bulk}^{\rm (2)} = \rho_1^{\rm (2)} = \alpha_{\rm 2}^*, \quad
J^{\rm (2)}=\alpha_{\rm 2}^*(1- \alpha_{\rm 2}^*),\quad
\rho_L^{\rm (2)} = \alpha_{\rm 2}^*(1-\alpha_{\rm 2}^*)/\beta_{\rm 2}^*.
\label{s2L}
\end{equation}
Substituting the expressions for $\rho_1^{\rm (2)}$ and $\rho_L^{\rm (2)}$ into
Eqs. (\ref{1}) and (\ref{3}), we find
\begin{equation}
\beta_{\rm 1}^* = 1-\alpha_{\rm 2}^*\left[1- q/(4\alpha_{\rm 3}^*)\right],
\quad \alpha_{\rm 3}^*= \alpha_{\rm 2}^*(1-\alpha_{\rm 2}^*)/\beta_{\rm 2}^* +
q/(4\beta_{\rm 1}^*).
\label{s13L}
\end{equation}
We have obtained a set of four nonlinear equations, see (\ref{s2eff}) and (\ref{s13L}), for
the four effective rates $\beta_{\rm 1}^*$, $\alpha_{\rm 2}^*$, $\beta_{\rm 2}^*$, and
$\alpha_{\rm 3}^*$. The solution depends on one free parameter, because
one of these equations is a consequence of the other three. From the first equation in
(\ref{s2eff}), we obtain the equation $\alpha_{\rm 2}^*= 1/(4\beta_{\rm 1}^*)-
J^{\rm sc} = 1/(4\beta_{\rm 1}^*)- 1/4 + \alpha_{\rm 2}^*(1-\alpha_{\rm 2}^*)$,
where we have used the relationship $J^{\rm sc} =1/4 - J^{\rm (2)}$ in combination with the expression
for $J^{\rm (2)}$ given in (\ref{s2L}). Hence, we find $\beta_{\rm 1}^* = 1/[1 + 4 (\alpha_{\rm 2}^*)^{2}]$.
After substitution of this expression into the first equation (\ref{s13L}), we solve the latter
for $\alpha_{\rm 3}^*$ and obtain $\alpha_{\rm 3}^*= q[1 + 4 (\alpha_{\rm 2}^*)^2]/[4(1-2\alpha_{\rm 2}^*)^2]$.
Finally, the second equation in (\ref{s2eff}) yields
$\beta_{\rm 2}^*=(1-2\alpha_{\rm 2}^*)^2/\{q[1 + 4 (\alpha_{\rm 2}^*)^2]\}$.
Now one can readily verify that the above expressions for $\beta_{\rm 1}^*$, $\alpha_{\rm 3}^*$ and
$\beta_{\rm 2}^*$ identically satisfy the second equation in (\ref{s13L}).

It remains to check the consistence of the results obtained with
the conditions for $(M,L,M)$ phase structure of the network. The
free parameter $\alpha_{\rm 2}^*$ has to satisfy the inequality
$\alpha_{\rm 2}^* <1/2$ which is necessary for the second segment
to be in low-density phase. The latter inequality implies
$\beta_{\rm 1}^* >1/2$ which, together with $\alpha >1/2$ ensures
that the first segment is in the maximum current phase. The
remaining condition $\alpha_{\rm 2}^* < \beta_{\rm 2}^*$ for the
second segment to be in the low-density phase leads to the cubic
inequality
\begin{equation}
4q (\alpha_{\rm 2}^*)^3 -4(\alpha_{\rm 2}^*)^2 +(4+q)\alpha_{\rm 2}^* -1 <0.
\label{cubic2L}
\end{equation}
This inequality has to be fulfilled simultaneously with the condition $\alpha_{\rm 3}^* >1/2$
for the third segment to be in the maximum current phase (given $\beta >1/2$), which implies
$(1-2\alpha_{\rm 2}^*)^2 < (q/2)[1+ 4(\alpha_{\rm 2}^*)^2].$
Therefore, the free parameter $\alpha_{\rm 2}^*$ has to obey the constraints
\begin{equation}
q \alpha_{\rm 2}^*[1+4(\alpha_{\rm 2}^*)^2] <(1-2\alpha_{\rm 2}^*)^2 < (q/2)[1+ 4(\alpha_{\rm 2}^*)^2]
\label{a3insL}
\end{equation}
which define a nonempty interval when $\alpha_{\rm 2}^* <1/2$. As a simple consequence, in the case of vanishing
probability of the shortcut, $q\rightarrow 0^-$, the free parameter
$\alpha_{\rm 2}^* \rightarrow 1/2^-$, which agrees with the result
$\alpha_{\rm 2}^* = \rho_{\rm bulk}=1/2$ for a single chain in the maximum current phase.

{\it Middle segment in the high-density phase.} --- In this case
the exact thermodynamic parameters of the second segment are
\begin{equation}
\rho_{\rm bulk}^{\rm (2)} = \rho_L^{\rm (2)} = 1-\beta_{\rm 2}^*, \quad
J^{\rm (2)}=\beta_{\rm 2}^*(1- \beta_{\rm 2}^*),\quad
\rho_1^{\rm (2)} = 1- \beta_{\rm 2}^*(1-\beta_{\rm 2}^*)/\alpha_{\rm 2}^*.
\label{s2H}
\end{equation}
Substituting the above expressions for $\rho_1^{\rm (2)}$ and $\rho_L^{\rm (2)}$ into
Eqs. (\ref{1}) and (\ref{3}), we find
\begin{equation}
\beta_{\rm 1}^* = q/(4\alpha_{\rm 3}^*) +
\left[1-q/(4\alpha_{\rm 3}^*)\right]\beta_{\rm 2}^*(1-\beta_{\rm 2}^*)/\alpha_{\rm 2}^*,
\quad \alpha_{\rm 3}^*= 1- \beta_{\rm 2}^* + q/(4\beta_{\rm 1}^*).
\label{s13H}
\end{equation}
Taking into account Eqs. (\ref{s2eff}), we have again a set of four nonlinear equations for
the four effective rates. We shall solve these equations
and show that one of them is a consequence of the other three. As in the previous case, one
of the effective rates will appear as a free parameter in the solution.
The second equation in (\ref{s2eff}) yields $\alpha_{\rm 3}^* = 1/(4 \beta_{\rm 2}^*)$.
Combining this result with the second equation in (\ref{s13H}), we express $\beta_{\rm 1}^*$ as
a function of $\beta_{\rm 2}^*$:
$\beta_{\rm 1}^* = q \beta_{\rm 2}^*/[1-4\beta_{\rm 2}^*(1-\beta_{\rm 2}^*)]$.
Taking into account that the current trough the shortcut is $J^{\rm sc} =1/4 - J^{\rm (2)}$, in view of the
present expression for the current $J^{\rm (2)}$, see Eq. (\ref{s2H}), we can rewrite the first
equation in (\ref{s2eff}) as
$\alpha_{\rm 2}^* = 1/(4 \beta_{\rm 1}^*) - 1/4 + \beta_{\rm 2}^*(1-\beta_{\rm 2}^*)$.
The substitution here of $\beta_{\rm 1}^*$ yields
$\alpha_{\rm 2}^* = \left[1/(q \beta_{\rm 2}^*) -1\right]\left[1/4 - \beta_{\rm 2}^*(1-\beta_{\rm 2}^*)
\right]$. One can readily check that the above expressions for $\alpha_{\rm 2}^*$, $\alpha_{\rm 3}^*$, and
$\beta_{\rm 1}^*$ satisfy the first equation in (\ref{s13H}) identically with respect to $\beta_{\rm 2}^* = 1-
\rho_L^{\rm (2)}$.

Finally, we check the conditions on the effective rates which
imply the phase structure $(M,H,M)$. The condition $\alpha_{\rm
2}^* > \beta_{\rm 2}^*$ leads to the cubic inequality
\begin{equation}
4q (\beta_{\rm 2}^*)^3 - 4 (\beta_{\rm 2}^*)^2 +(4 +q)\beta_{\rm 2}^* -1 >0,
\label{cubeH}
\end{equation}
which, together with $\beta_{\rm 2}^* < 1/2$ ensures the high-density phase of the second
segment. The first segment is in the maximum current phase when $\alpha >1/2$ and $\beta_{\rm 1}^* >1/2$.
The second condition leads to the inequality
$q \beta_{\rm 2}^* > (1/2)[1- 4 \beta_{\rm 2}^*(1-\beta_{\rm 2}^*)]$.
The right-hand side being non-negative, $q\rightarrow 0$ implies $\beta_{\rm 2}^* \rightarrow 1/2$,
hence $J^{\rm (2)}\rightarrow 1/4$ and $J^{\rm sc}\rightarrow 0$. The analysis of this quadratic
inequality is equivalent to $\beta_{\rm 2}^* < 1/2 + q/4 - (1/4)\sqrt{q(4+ q)}$,
the right-hand side of which is less than 1/2. The condition $\alpha_{\rm 3}^* >1/2$ for
the third segment to be in the maximum current phase (given $\beta >1/2$) is satisfied whenever $\beta_{\rm 2}^* <1/2$.

\begin{figure}[t]
  \includegraphics[width=100mm]{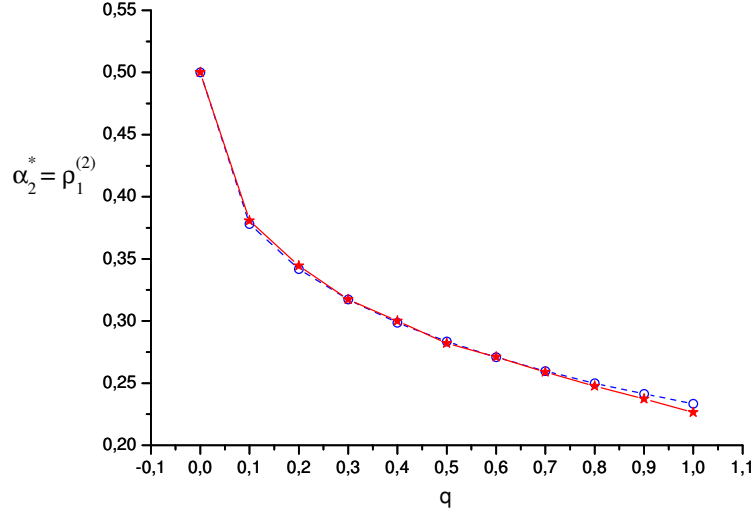}
  \label{Fig1}
  \caption{(Color online) Comparison of the numerically evaluated $a_2^* = \rho_1^{(2)}$, shown by red
   stars, with the values of the appropriate root of the cubic equation (\ref{cubic}), shown by blue
   circles, for ten different values of the rate $q$.}
   \label{Fig1}
\end{figure}

{\it Middle segment on the coexistence line.} --- The second
segment can exist in a low- or high-density phase, depending on
whether $\alpha_{\rm 2}^* > \beta_{\rm 2}^*$ or $\alpha_{\rm 2}^*
< \beta_{\rm 2}^*$, respectively. Naturally, we expect the
coexistence phase (shock phase) to take place at a common point
 in the closure of the above open sets, i.e., when the rates $\alpha_{\rm 2}^* = \beta_{\rm 2}^*$
 coincide with an
appropriate root of the cubic equation given by an equality sign in expressions (\ref{cubeH}) and
(\ref{cubic2L}). To prove this, we set $\alpha_{\rm 2}^* = \beta_{\rm 2}^*$ and assume the exact in the
thermodynamic limit values of the current and the local densities at the endpoints of the
second segment in the coexistence phase,
\begin{equation}
J^{\rm (2)}=\alpha_{\rm 2}^*(1-\alpha_{\rm 2}^*)=\beta_{\rm 2}^*(1-\beta_{\rm 2}^*), \quad
\rho_{1}^{\rm (2)} = \alpha_{\rm 2}^* = \rho_{-}(J^{\rm (2)}),\quad
\rho_L^{\rm (2)} = 1-\alpha_{\rm 2}^* = \rho_{+}(J^{\rm (2)}).
\label{s2C}
\end{equation}
Here $\rho_{\pm}(J) =(1\pm \sqrt{1-4J})/2$
are the bulk densities in the high- and low-density phase, respectively, of a single chain with a
current $J$. Substituting the above expressions for $\rho_1^{\rm (2)}$ and $\rho_L^{\rm (2)}$ into
Eqs. (\ref{1}) and (\ref{3}), we obtain
\begin{equation}
\beta_{\rm 1}^* = \rho_{+}(J^{\rm (2)}) + q\rho_{+}(J^{\rm (2)})/(4\alpha_{\rm 3}^*), \quad
\alpha_{\rm 3}^*= \rho_{+}(J^{\rm (2)}) + q/(4\beta_{\rm 1}^*).
\label{s13C}
\end{equation}

\begin{figure} [!t]
  \includegraphics[width=100mm]{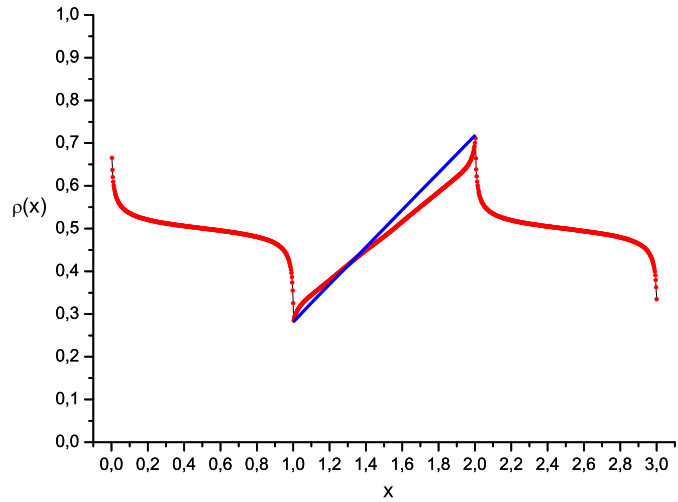}
  \label{Fig2}
\caption{(Color online) Local density profile at $\alpha = \beta =0.75$ and $q=0.5$ shown by red
   stars as a function of the normalized coordinate $x=i/L$, where $i= (s-1)L +1,(s-1)L +2,\dots,sL$
   labels the sites in the segment $s$, $s=1,2,3$. The shape of the density profile in the first and
   third segments is    typical for the MC phase, while that in the second segment closely resembles
   the linear dependence
   with the distance characteristic of the coexistence phase with completely delocalized domain wall.
   The predictions of the domain wall theory are shown by a blue line.}
\end{figure}

Inserting in the first equation $1/(4\alpha_{\rm 3}^*)$ expressed from the second equation in (\ref{s2eff})
with $\beta_{\rm 2}^* = \alpha_{\rm 2}^*$, and replacing $\rho_{+}(J^{\rm (2)})$ and $\rho_{-}(J^{\rm (2)})$
by $1-\alpha_{\rm 2}^*$ and $\alpha_{\rm 2}^*$, respectively, we obtain
$\beta_{\rm 1}^* = 1 - \alpha_{\rm 2}^* + q(\alpha_{\rm 2}^*)^2$.
On the other hand, dividing both sides of the first equation in (\ref{s13C}) by $4\beta_{\rm 1}^*$,
and using $J^{\rm sc}= 1/4 - J^{\rm (2)}$, we arrive at
$1/4 = \rho_{+}(J^{\rm (2)})/(4\beta_{\rm 1}^*) + \rho_{-}(J^{\rm (2)})\left(1/4 - J^{\rm (2)}\right)$.
Solving the above for $\beta_{\rm 1}^*$, and using that $J^{\rm (2)}= \rho_{+}(J^{\rm (2)})\rho_{-}(J^{\rm (2)})=
\rho_{+}(J^{\rm (2)})\alpha_{\rm 2}^*$, we obtain $\beta_{\rm 1}^* = [1+ 4(\alpha_{\rm 2}^*)^2]^{-1}$.
Clearly, $\beta_{\rm 1}^* >1/2$ when $\alpha_{\rm 2}^* <1/2$. The equality the right-hand sides of the two above
derived expressions for $\beta_{\rm 1}^*$ leads to the equation
\begin{equation}
4q (\alpha_{\rm 2}^*)^3 -4(\alpha_{\rm 2}^*)^2 +(4+q)\alpha_{\rm 2}^* -1 = 0.
\label{cubic}
\end{equation}

Unexpectedly, the value of $\alpha_{\rm 2}^* = \beta_{\rm 2}^*$ is
determined by a root of cubic equation, which is singular at
$q=0$. More precisely, these effective rates are given as function
of $q$, $0\leq q \leq 1$, by the real root of Eq. (\ref{cubic})
which is less than 1/2 and tends to 1/2 from below as $q
\rightarrow 0^+$. A comparison of the values of $\alpha_{\rm 2}^*$
given by the appropriate root of (\ref{cubic}) and $\rho_{1}^{\rm
(2)}$ evaluated by computer simulations are shown in Fig. 1. From
the second equation in (\ref{s2eff}) at $\beta_{\rm 2}^* =
\alpha_{\rm 2}^*$ we have $\alpha_{\rm 3}^* = 1/(4\alpha_{\rm
2}^*)$, so that $\alpha_{\rm 2}^* <1/2$ directly implies
$\alpha_{\rm 3}^* >1/2$.

{\it Predictions of the domain wall theory.} --- An open chain
with stationary current $J<1/4$ in the thermodynamic limit can be
found in two phases: with low density $\rho_{-}(J)$ and high
density  $\rho_{+}(J)$. According to the domain wall theory
\cite{DW}, on the coexistence line the two phases are
separated by a completely delocalized domain wall. As a result,
the averaged density profile is linear, changing its value from
$\rho_{-}(J)$ at the left end of the chain to $\rho_{+}(J)$ at its
right end. This prediction is compared to numerical simulation
data in Fig.~2 for external rates $\alpha =\beta = 0.75$, length
of each segment $L =400$, and $q=0.5$. The data was averaged over
100 runs of length $2^{23}$ attempted moves each. One sees a very
good agreement between the theoretical prediction $\rho_1^{(2)}=
\rho_{-}(J^{(2)})\simeq 0.282$ and the simulation result
$\rho_1^{(2)}\simeq 0.286$. Fairly good is also the agreement at
the high-density end, between the theoretical prediction
$\rho_{400}^{(2)}= \rho_{+}(J^{(2)})\simeq 0.718$ and the
simulation result $\rho_{400}^{(2)}\simeq 0.701$.

\begin{figure}[!t]
\includegraphics[width=100mm]{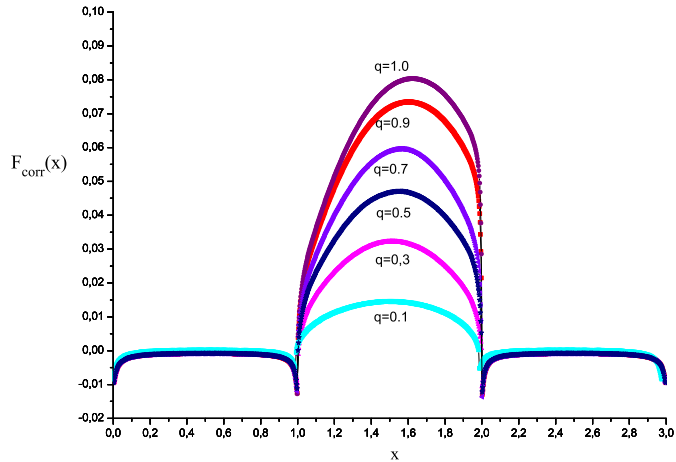} \caption{
   (Color online)  Position dependence of the nearest-neighbor correlations
    along the network at different values of $q$.
    The normalized coordinate $x=i/L$ is the same as in Fig. 2.}
\end{figure}
\begin{figure}[!t]
\includegraphics[width=100mm]{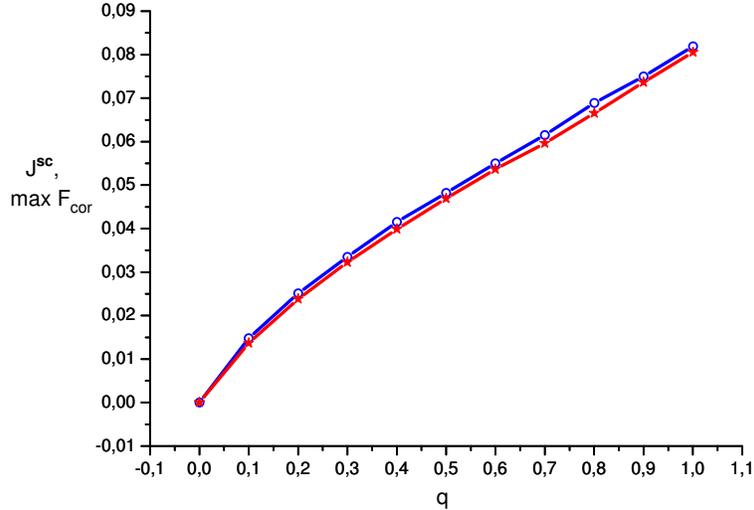} \caption{
 (Color online) Comparison between the numerically estimated maximum value
   of the nearest-neighbor correlations in the second segment $\max_x F_{\rm cor}(x)$,
   shown by red stars connected with a red line, and the current $J^{\rm sc}$ through
   the shortcut, shown by blue circles connected with a blue line, at different values of
   the parameter $q$. }
\label{Fig4}
\end{figure}

Another important prediction of the domain wall theory is the
parabolic shape of the nearest-neighbor correlations $F_{\rm
cor}(x)= \langle \tau_i^{\rm (2)}\tau_{i+1}^{\rm (2)}\rangle -
\langle \tau_i^{\rm (2)}\rangle \langle\tau_{i+1}^{\rm
(2)}\rangle$ as a function of the normalized distance $x= i/L$.
The simulation results for all $q$ show almost vanishing correlations
in the bulk of the first  and third segments and a parabolic-like shape
in the second segment, see Fig. 3. In the latter case the noticeable tilt of the 'parabolas'
 to the right when $q \geq 0.7$ may be due to the different value of the correlations
$G^{(s,s+1)}$ between the segments $s$ and $s+1$, $s=1,2$. For example, we have
numerically evaluated $G^{(1,2)}\simeq -0.0073$ at both $q=0.3$ and $q=1.0$,
while $G^{(2,3)}\simeq -0.0023$ at $q=0.3$ but $G^{(2,3)}\simeq 0.0095$ at $q=1.0$.
Theoretically, the maximum value of $F_{\rm cor}(x)$ is reached at the midpoint of the chain and
equals $$\max_x F_{\rm cor}(x) = [\rho_{+}(J^{(2)})-\rho_{-}(J^{(2)})]^2/4 = 1/4 - J^{(2)} =
J^{\rm sc}.$$ The validity of this prediction of the domain wall theory is illustrated in Fig. 4.

{\it Conclusions.} --- The model predicts that a shortcut in the bulk of a road carrying
maximum stationary current inevitably causes traffic jams characteristic of a shock phase
with completely delocalized domain wall. The main parameters of the average density profile
and the nearest-neighbor correlations in the shunted segment
are governed by a cubic equation with coefficients simply depending on the
probability of choosing the shortcut.

{\it Acknowledgements.} --- N.B. gratefully acknowledges support from a grant of the Representative
Plenipotentiary of Bulgaria to the Joint Institute for Nuclear Research in Dubna. N.P. thanks Roumen
Anguelov and Jean Lubuma for their hospitality at the University of Pretoria, where a part of this
work was also carried out.

\end{document}